# Alternatives to Classical Option Pricing

W. Brent Lindquist*, Svetlozar T. Rachev

Department of Mathematics & Statistics, Texas Tech University, Lubbock, TX USA

*Corresponding author, brent.lindquist@ttu.edu

**Abstract.** We develop two alternate approaches to arbitrage-free, market-complete, option pricing. The first approach requires no riskless asset. We develop the general framework for this approach and illustrate it with two specific examples. The second approach does use a riskless asset. However, by ensuring equality between real-world and risk-neutral price-change probabilities, the second approach enables the computation of risk-neutral option prices utilizing expectations under the natural world probability $\mathbb{P}$. This produces the same option prices as the classical approach in which prices are computed under the risk neutral measure $\mathbb{Q}$. The second approach and the two specific examples of the first approach require the introduction of new, marketable asset types, specifically perpetual derivatives of a stock, and a stock whose cumulative return (rather than price) is deflated.

## 1. Introduction

The classical development of the theory of dynamic asset and derivative pricing rests on the existence of a riskless asset providing a predictable return. Generally, these are assumed to be rates set by government debt or rates governing interbank loans. Regarding the former, past Under Secretary of the U.S. Treasury, Peter Fisher (2013), has stated "The idea of risk-free sovereign bonds is best thought of as an oxymoron or as an anomaly of recent history. It is not a useful, necessary or an enduring feature of the financial landscape." Regarding the latter, the Libor scandal in 2012 revealed that such rates are susceptible to manipulation.

In an early investigation of the role of a riskless asset, Black (1972) confirmed that the capital asset model holds in the absence of a riskless rate. Equilibrium models with riskless assets have been considered (Nielsen, 1990; Allingham, 1991; Konno and Shirakawa, 1995; Sun and Yang, 2003). More recent papers have focused on the challenges involved when there are riskless asset shortages (Eggertsson and Krugman, 2012; Gourinchas and Jeanne, 2012; Caballero and Farhi, 2013; Aoki, Nakajima and Nikolov, 2014). Rachev, Stoyanov and Fabozzi (2017) have derived Black-Scholes-Merton (BSM) option pricing equations for markets containing only risky assets. They considered various price dynamics, including continuous and jump diffusions, stochastic volatility, and geometric fractional Brownian and Rosenblatt motions.

In this paper, we investigate two alternative approaches to option pricing. Approach one considers option pricing in a market consisting solely of risky assets. While the work of Rachev, Stoyanov and Fabozzi addressed the existence of a riskless rate based only upon risky assets, in Section 2, using the continuous diffusion dynamics for the risky asset prices, we address the practical aspects of implementing such a rate and establish a general framework for implementing the option pricing. Specifically, we consider a market consisting of two risky assets (stocks) and a European contingent claim (option) whose underlying consists of a portfolio composed of the two stocks. We assume that the price processes of the two stocks are driven by the same Brownian motion.[1] In the absence of a riskless asset, one of the stock prices serves as the numéraire governing the relationship between the risk-neutral and natural world probabilities. In this market we develop the stochastic PDE governing the option price (it is sufficient to work with call options). We develop the Feynman-Kac solution for the option price; the solution of which involves a parameter whose form is not available analytically. However, we explicitly verify that the constant coefficient version of the Feynman-Kac solution does solve the stochastic PDE. To provide a practical method of implementing option pricing in this market for time-varying parameters, we develop the corresponding discrete binomial model.

The classical pricing models, whether continuous (e.g. BSM) or discrete (e.g. binomial), replace the natural world price-change probabilities with risk-neutral (riskless) probabilities. In Section 3, we develop a second approach to option pricing which does include a riskless asset but considers a risky asset whose cumulative return process is deflated (as opposed to deflating its price process). As a result, the risk-neutral probabilities coincide with the natural world probabilities, enabling the development of risk-neutral option prices in an arbitrage-free, complete market, using expectations that rely solely on the natural world probability $\mathbb{P}$.

[1] In theory this can be extended to the case of $n$ risky assets driven by $n-1$ Brownian motions.

In Section 4, we consider two specific examples to approach one. In the first example the numéraire asset has the deflated cumulative return process introduced in Section II. In the second example, the numéraire asset is a perpetual derivative of the first.

A discussion of the implications and benefits of these alternate approaches is provided in Section 5. In particular, these approaches resolve two critical option price discontinuity puzzles (Hu, Shirvani, Stoyanov, et al., 2020).

## 2. Approach One: General Framework

We consider two, perfectly correlated, risky assets (stocks) $\mathcal{S}$ and $\mathcal{Z}$ and an option $\mathcal{C}$. The price processes $S(t), Z(t)$ for the two stocks are given by[2]

$$dS(t) = S(t)\big(\mu(t)dt + \sigma(t)dW(t)\big), \qquad dZ(t) = Z(t)(\tilde{\mu}(t)dt + \tilde{\sigma}(t)dW(t)), \tag{1}$$

where $W(t)$ is a Wiener process that generates a stochastic basis $(\Omega, \mathcal{F}, \{\mathcal{F}(t)\}_{t\geq 0}, P)$ representing the natural world on which price processes are defined.

### *A. Derivation and Solution of the Stochastic PDE for the Option*

The price process $C(t, S(t), Z(t))$ for the option obeys the Itô stochastic differential equation

$$\begin{gathered} dC = \big[C_t + C_\mu + C_{dd}\big]dt + C_\sigma dW(t), \\ C_\mu \equiv \mu(t)S(t)C_S + \tilde{\mu}(t)Z(t)C_Z, \quad C_\sigma \equiv \sigma(t)S(t)C_S + \tilde{\sigma}(t)Z(t)C_Z, \\ C_{dd} \equiv \frac{1}{2}(\sigma(t)^2S(t)^2C_{SS} + 2\sigma(t)\tilde{\sigma}(t)S(t)Z(t)C_{ZS} + \tilde{\sigma}(t)^2Z(t)^2C_{ZZ}) . \end{gathered} \tag{2}$$

Following the BSM formulation, the usual conditions

$$\pi(t, S, Z) = 0, \qquad d\pi(t, S, Z) = 0, \tag{3}$$

applied to the self-financing portfolio $\pi(t, S, Z) = a(t)S(t) + b(t)Z(t) - C(t, S(t), Z(t))$ produces the partial differential equation for the option price,

$$\bar{r}(t)C = C_t + \bar{r}(t)C_d + C_{dd}\, , \qquad C_d \equiv S(t)C_S + Z(t)C_Z\, , \tag{4}$$

where

$$\bar{r}(t) = \frac{\mu(t)\tilde{\sigma}(t) - \tilde{\mu}(t)\sigma(t)}{\Delta\sigma(t)}\, , \qquad \Delta\sigma(t) = \tilde{\sigma}(t) - \sigma(t)\, . \tag{5}$$

[2] There is a convention of denoting a time dependent stochastic process by $X_t$. In Section 2A we employ alphabetic subscripts to indicate terms involving partial derivatives. Thus $X(t,\cdot)$ denotes a process having explicit dependence on time, while $X_t$ denotes its partial derivative with respect to time.

We shall refer to (4) as the Lindquist-Rachev (LR) PDE.[3]

We have found no direct method of solving (4) analytically for a European call option. We therefore develop a solution for the option price using the Feynman-Kac formula and then verify that it satisfies (4). To do so in the absence of any riskless asset (i.e. a bond), we utilize the price of stock $\mathcal{S}$ as the numéraire governing the relationship between the risk-neutral and natural world probabilities. Under this numéraire, $\hat{S}(t) \equiv S(t)/S(t) = 1$ and, via Itô's formula, $\hat{Z}(S,Z) \equiv Z(t)/S(t)$ obeys

$$d\hat{Z}(S,Z) = \hat{Z}(S,Z)\left([\Delta\mu(t) - \sigma(t)\Delta\sigma(t)]dt + \Delta\sigma(t)dW(t)\right), \tag{6}$$

where $\Delta\mu(t) = \tilde{\mu}(t) - \mu(t)$. Using Girsanov's theorem to change probability measures via $dW(t) = dW^Q(t) - \theta(t)dt$ with a market price of risk $\theta(t) = \Delta\mu(t)/\Delta\sigma(t) - \sigma(t)$, under the risk-neutral probability measure $Q$ the price processes obey

$$\begin{aligned} d\hat{Z}^Q(S,Z) &= \hat{Z}^Q(S,Z)\,\Delta\sigma(t)dW^Q(t), \\ dS^Q(t) &= S^Q(t)([\bar{r}(t) + \sigma(t)^2]dt + \sigma(t)dW^Q(t)), \\ dZ^Q(t) &= Z^Q(t)([\bar{r}(t) + \sigma(t)\tilde{\sigma}(t)]dt + \tilde{\sigma}(t)dW^Q(t)). \end{aligned} \tag{7}$$

Under measure $Q$, $\hat{Z}^Q$ is a martingale. We also note that, under the natural world probability $P$, the Sharpe ratios for stocks $\mathcal{S}$ and $\mathcal{Z}$, whose prices are driven by the same Wiener process, must be equal,

$$\frac{\mu(t) - r^P(t)}{\sigma(t)} = \frac{\tilde{\mu}(t) - r^P(t)}{\tilde{\sigma}(t)} \;\rightarrow\; r^P(t) = \frac{\mu(t)\tilde{\sigma}(t) - \tilde{\mu}(t)\sigma(t)}{\Delta\sigma(t)} = \bar{r}(t)\,, \tag{8a}$$

while the same equality of Sharpe ratios must hold under the risk-neutral probability $Q$,

$$\frac{\bar{r}(t) + \sigma(t)^2 - r^Q(t)}{\sigma(t)} = \frac{\bar{r}(t) + \sigma(t)\tilde{\sigma}(t) - r^Q(t)}{\tilde{\sigma}(t)} \;\rightarrow\; r^Q(t) = \bar{r}(t)\,, \tag{8b}$$

confirming that the risk-free rate $\bar{r}(t)$ is identical under probabilities $P$ and $Q$.

Note that the riskless rate $\bar{r}(t)$ is a two-asset example of the general shadow riskless rate developed in the aforementioned work of Rachev, Stoyanov and Fabozzi (2017). In general, $\bar{r}(t)$ will be based upon a large market of $n$ assets and will be given by the ratio of two determinants, each of size $n \times n$. In practice, the shadow riskless rate based upon an entire market of $n$ assets will be used in the option pricing.

[3] While we have formulated the LR PDE for an option, it holds for any process $f(t, S(t), Z(t))$ with $S(t), Z(t))$ given by (1). Solution of the PDE depends, as usual, on boundary and initial conditions.

From (7), the price processes $S^Q(t)$ and $Z^Q(t)$ are

$$S^Q(T)|\big(S^Q(t)=S(t)\big) = S(t)\exp\left\{\int_t^T\left[\bar{r}(s)+\frac{\sigma(s)^2}{2}\right]ds+\int_t^T\sigma(s)dW^Q(s)\right\},$$

$$Z^Q(T)|(Z^Q(t)=Z(t)) \qquad (9)$$

$$= Z(t)\exp\left\{\int_t^T\left[\bar{r}(s)+\sigma(s)\tilde{\sigma}(s)-\frac{\tilde{\sigma}(s)^2}{2}\right]ds+\int_t^T\tilde{\sigma}(s)dW^Q(s)\right\}.$$

We denote the "per-share" risk-neutral value of the portfolio $\{\mathcal{S},\mathcal{Z}\}$ at time $T$ by

$$\eta S^Q(T)+(1-\eta)Z^Q(T), \qquad (10)$$

where $\eta$ is the fraction of total portfolio shares held in stock $\mathcal{S}$.[4] With a time to maturity $T$, a per-share strike price $K$, and using $S^Q(t)$ as the price numéraire, the Feynman-Kac formula for the price of the call option with this portfolio as the underlying security is (Duffie, 2001, 116)

$$C(t,S(t),Z(t);K,T)=E_t^Q\left[\frac{S(t)}{S^Q(T)}\max(0,\ \eta S^Q(T)+(1-\eta)Z^Q(T)-K)\right]$$

$$=E_t^Q\Bigg[\max\Bigg(0,\ \eta\, S(t)$$

$$+(1-\eta)Z(t)\exp\left\{\int_t^T\left[-\frac{\big(\sigma(s)-\tilde{\sigma}(s)\big)^2}{2}\right]ds-\int_t^T\big(\sigma(s)-\tilde{\sigma}(s)\big)dW^Q(s)\right\}$$

$$-K\exp\left\{-\int_t^T\left[\bar{r}(s)+\frac{\sigma(s)^2}{2}\right]ds-\int_t^T\sigma(s)dW^Q(s)\right\}\ \Bigg)\Bigg]. \qquad (11)$$

Equation (11) is analytically intractable. To provide further analysis, we assume time-independent risk-free rate, $\bar{r}(t)=\bar{r}$, and volatilities, $\sigma(t)=\sigma$, $\tilde{\sigma}(t)=\tilde{\sigma}$. Then the price processes (9) are log-normal,

$$\ln[S^Q(T)\,|(S^Q(t)=S(t))]\sim\mathcal{N}\big(\ln[S(t)]+m(t)+w(t)^2/2\,,w(t)\big),$$

$$\ln[Z^Q(T)|(Z^Q(t)=Z(t))]\sim\mathcal{N}\big(\ln[Z(t)]+m(t)+w(t)\tilde{w}(t)-\tilde{w}(t)^2/2\,,\tilde{w}(t)\big), \qquad (12)$$

where $m(t)=\bar{r}(T-t)$, $w(t)=\sigma\sqrt{T-t}$, $\tilde{w}(t)=\tilde{\sigma}\sqrt{T-t}$, and $\sim\mathcal{N}(\cdot,\cdot)$ denotes a normal random variable. The call option price (11) is now

[4] A more complete approach would be to designate the per-share portfolio value at time $T$ by $\eta(T)S^Q(T)+(1-\eta(T))Z^Q(T)$ and use Merton's approach (Duffie, 2001, chapter 9B) to optimize the value for $\eta(T)$ given prices at time $t$.

$$C(t,S(t),Z(t);K,T) = \int_{y^*}^{\infty} \big( \eta S(t) + (1-\eta)Z(t)\exp[-\Delta w^2(t)/2 - \Delta w(t)y]$$
$$-K\exp[-m(t) - w^2(t)/2 - w(t)y] \big)\,\varphi(y)dy\,. \quad (13)$$

where: $\Delta w(t) \equiv w(t) - \widetilde{w}(t)$; $\varphi(y) = \left(\sqrt{2\pi}\right)^{-1}\exp\,(-y^2/2)$ is the standard normal density; and $y^*$ satisfies the nonlinear equation

$$\eta S(t)\exp[m(t) + w^2(t)/2 + w(t)y^*]$$
$$+(1-\eta)Z(t)\exp[m(t) + w(t)\widetilde{w}(t) - \widetilde{w}^2(t)/2 + \widetilde{w}(t)y^*] = K. \quad (14)$$

Using (12), the solution (13) can be written

$$C(t,S(t),Z(t);K,T)$$
$$= \eta S(t)\Phi(d) + (1-\eta)Z(t)\,\Phi(d - \Delta w(t)) - Ke^{-m(t)}\Phi(d - w(t)), \quad (15)$$

where $d \equiv -y^*$ and $\Phi(x) = \int_{-\infty}^{x}\varphi(y)dy$ is the standard normal cumulative distribution function.

To verify that (15) is a solution of the LR PDE, we require certain partial derivatives of $y^*$ with respect to $t, S, Z$. In the absence of an analytic solution for $y^*$, it is possible to implicitly determine the required partial derivatives of $y^*$ from (14). In the appendix, we determine these partial derivatives and show that the solution (15) satisfies the LR PDE.

*B. Binomial Tree Solution of the Option Price*

In the absence of a fully analytic solution, either to the LR PDE or to (15), we consider a binomial tree solution. For simplicity, we assume a constant time increment for the lattice; $t_k = k\Delta_n,\ k = 0, \dots, n;\ \ \Delta_n = T/n$, with $T > 0$ being a terminal time. Consider a sequence $\xi_1^{(\Delta)}, \dots, \xi_n^{(\Delta)}$ of independent Bernoulli random variables with $\mathbb{P}^{(\Delta)}\left(\xi_k^{(\Delta)} = 1\right) = 1 - \mathbb{P}^{(\Delta)}\left(\xi_k^{(\Delta)} = 0\right) = p_{t_k}^{(\Delta)}$, $k = 1, \dots, n$.[5] On the tree, the prices of the stocks $\mathcal{S}$ and $\mathcal{Z}$ follow the discrete process[6]

$$S_{t_{k+1}}^{(\Delta)} = \begin{cases} S_{t_{k+1}}^{(\Delta,\mathrm{u})} = (1 + U_{t_{k+1}}^{(\Delta)})S_{t_k}^{(\Delta)} & \text{w.p.} \quad p_{t_{k+1}}^{(\Delta)}, \\ S_{t_{k+1}}^{(\Delta,\mathrm{d})} = (1 + D_{t_{k+1}}^{(\Delta)})S_{t_k}^{(\Delta)} & \text{w.p.} \quad 1 - p_{t_{k+1}}^{(\Delta)}, \end{cases} \quad (16a)$$

$$Z_{t_{k+1}}^{(\Delta)} = \begin{cases} Z_{t_{k+1}}^{(\Delta,\mathrm{u})} = (1 + \widetilde{U}_{t_{k+1}}^{(\Delta)})Z_{t_k}^{(\Delta)} & \text{w.p.} \quad p_{t_{k+1}}^{(\Delta)}, \\ Z_{t_{k+1}}^{(\Delta,\mathrm{d})} = (1 + \widetilde{D}_{t_{k+1}}^{(\Delta)})Z_{t_k}^{(\Delta)} & \text{w.p.} \quad 1 - p_{t_{k+1}}^{(\Delta)}, \end{cases} \quad (16b)$$

[5] We use the superscript $(\Delta)$ to designate discrete-time price processes.
[6] As the model is now discrete, there is no need to reserve notation for partial derivatives, and we resort to a customary notation where alphabetic subscripting indicates time dependence.

with $U_{t_k}^{(\Delta)} > D_{t_k}^{(\Delta)} > 0$, $\widetilde{U}_{t_k}^{(\Delta)} > \widetilde{D}_{t_k}^{(\Delta)} > 0$, $S_{t_0}^{(\Delta)} \equiv S_0 > 0$, and $Z_{t_0}^{(\Delta)} \equiv Z_0 > 0$. For $t \in [t_k, t_{k+1})$, $k = 0, 1, \dots, n-1$, we set $S_t^{(\Delta)} = S_{t_k}^{(\Delta)}$ with $S_T^{(\Delta)} = S_{t_n}^{(\Delta)}$, and $Z_t^{(\Delta)} = Z_{t_k}^{(\Delta)}$ with $Z_T^{(\Delta)} = Z_{t_n}^{(\Delta)}$, establishing piecewise smooth pricing processes on the tree. The (independent) arithmetic returns are given by

$$r_{t_{k+1}}^{(S,\Delta)} = \frac{S_{t_{k+1}}^{(\Delta)} - S_{t_k}^{(\Delta)}}{S_{t_k}^{(\Delta)}} = \begin{Bmatrix} r_{t_{k+1}}^{(S,\Delta,\mathrm{u})} = U_{t_{k+1}}^{(\Delta)} & \text{w.p.} & p_{t_{k+1}}^{(\Delta)} \\ r_{t_{k+1}}^{(S,\Delta,\mathrm{d})} = D_{t_{k+1}}^{(\Delta)} & \text{w.p.} & 1 - p_{t_{k+1}}^{(\Delta)} \end{Bmatrix}, \tag{17a}$$

$$r_{t_{k+1}}^{(Z,\Delta)} = \frac{Z_{t_{k+1}}^{(\Delta)} - Z_{t_k}^{(\Delta)}}{Z_{t_k}^{(\Delta)}} = \begin{Bmatrix} r_{t_{k+1}}^{(Z,\Delta,\mathrm{u})} = \widetilde{U}_{t_{k+1}}^{(\Delta)} & \text{w.p.} & p_{t_{k+1}}^{(\Delta)} \\ r_{t_{k+1}}^{(Z,\Delta,\mathrm{d})} = \widetilde{D}_{t_{k+1}}^{(\Delta)} & \text{w.p.} & 1 - p_{t_{k+1}}^{(\Delta)} \end{Bmatrix}, \tag{17b}$$

$$k = 0, \dots, n-1, \qquad r_{t_0}^{(S,\Delta)} = r_{t_0}^{(Z,\Delta)} \equiv r_0 = 0\,.$$

The stocks $\mathcal{S}$ and $\mathcal{Z}$ constitute the replicating portfolio for the option price. The standard replicating portfolio conditions give the binomial tree relation for the option price,

$$C_{t_k}^{(\Delta)} = \frac{q_{t_{k+1}}^{(\Delta)} C_{t_{k+1}}^{(\Delta,\mathrm{u})} + (1 - q_{t_{k+1}}^{(\Delta)}) C_{t_{k+1}}^{(\Delta,\mathrm{d})}}{R_{t_{k+1}}^{(\Delta)}}, \tag{18}$$

where the risk-neutral probability is

$$q_{t_{k+1}}^{(\Delta)} = \frac{\Delta D_{t_{k+1}}}{\Delta D_{t_{k+1}} - \Delta U_{t_{k+1}}}, \qquad \Delta U_{t_{k+1}} \equiv \widetilde{U}_{t_{k+1}}^{(\Delta)} - U_{t_{k+1}}^{(\Delta)}, \qquad \Delta D_{t_{k+1}} \equiv \widetilde{D}_{t_{k+1}}^{(\Delta)} - D_{t_{k+1}}^{(\Delta)}, \tag{19}$$

and the cumulative return $R_{t_{k+1}}^{(\Delta)}$ over the interval $k\Delta_n \rightarrow (k+1)\Delta_n$ is

$$R_{t_{k+1}}^{(\Delta)} = \frac{(1 + U_{t_{k+1}}^{(\Delta)})(1 + \widetilde{D}_{t_{k+1}}^{(\Delta)}) - (1 + \widetilde{U}_{t_{k+1}}^{(\Delta)})(1 + D_{t_{k+1}}^{(\Delta)})}{\Delta D_{k+1} - \Delta U_{k+1}}. \tag{20}$$

With the first two moments of the arithmetic returns of $S_{t_{k+1}}^{(\Delta)}$ and $Z_{t_{k+1}}^{(\Delta)}$ given by $E\left[r_{t_{k+1}}^{(S,\Delta)}\right] = \mu_{t_k}^{(\Delta)}\Delta_n$, $\mathrm{Var}\left[r_{t_{k+1}}^{(S,\Delta)}\right] = \tilde{\sigma}_{t_k}^{(\Delta)^2}\Delta_n$, $E\left[r_{t_{k+1}}^{(Z,\Delta)}\right] = \tilde{\mu}_{t_k}^{(\Delta)}\Delta_n$, and $\mathrm{Var}\left[r_{t_{k+1}}^{(Z,\Delta)}\right] = \tilde{\sigma}_{t_k}^{(\Delta)^2}\Delta_n$, then

$$U_{t_{k+1}}^{(\Delta)} = \mu_{t_k}^{(\Delta)}\Delta_n + \sigma_{t_k}^{(\Delta)}\sqrt{\frac{1 - p_{t_{k+1}}^{(\Delta)}}{p_{t_{k+1}}^{(\Delta)}}\Delta_n}\,, \tag{21a}$$

$$D_{t_{k+1}}^{(\Delta)} = \mu_{t_k}^{(\Delta)}\Delta_n - \sigma_{t_k}^{(\Delta)}\sqrt{\frac{p_{t_{k+1}}^{(\Delta)}}{1 - p_{t_{k+1}}^{(\Delta)}}\Delta_n}\,, \tag{21b}$$

$$\widetilde{U}_{t_{k+1}}^{(\Delta)} = \tilde{\mu}_{t_k}^{(\Delta)}\Delta_n + \tilde{\sigma}_{t_k}^{(\Delta)}\sqrt{\frac{1-p_{t_{k+1}}^{(\Delta)}}{p_{t_{k+1}}^{(\Delta)}}\Delta_n}, \qquad \widetilde{D}_{t_{k+1}}^{(\Delta)} = \tilde{\mu}_{t_k}^{(\Delta)}\Delta_n - \tilde{\sigma}_{t_k}^{(\Delta)}\sqrt{\frac{p_{t_{k+1}}^{(\Delta)}}{1-p_{t_{k+1}}^{(\Delta)}}\Delta_n}.$$

Equations (19) and (20) then become

$$q_{t_{k+1}}^{(\Delta)} = p_{t_{k+1}}^{(\Delta)} - \frac{\Delta\mu_{t_k}}{\Delta\sigma_{t_k}}\sqrt{p_{t_{k+1}}^{(\Delta)}\left(1-p_{t_{k+1}}^{(\Delta)}\right)\Delta_n}, \qquad R_{t_{k+1}}^{(\Delta)} = 1 + \bar{r}_{t_k}^{(\Delta)}\Delta_n,$$
$$\Delta\mu_{t_k} = \tilde{\mu}_{t_k}^{(\Delta)} - \mu_{t_k}^{(\Delta)}, \qquad \Delta\sigma_{t_k} = \tilde{\sigma}_{t_k}^{(\Delta)} - \sigma_{t_k}^{(\Delta)}, \qquad \bar{r}_{t_k}^{(\Delta)} = \frac{\mu_{t_k}^{(\Delta)}\tilde{\sigma}_{t_k}^{(\Delta)} - \tilde{\mu}_{t_k}^{(\Delta)}\sigma_{t_k}^{(\Delta)}}{\Delta\sigma_{t_k}}. \tag{22}$$

The terms $\Delta\mu_{t_k}$, $\Delta\sigma_{t_k}$ and $\bar{r}_{t_k}^{(\Delta)}$ are the discrete analogs of $\Delta\mu(t), \Delta\sigma(t)$, and $\bar{r}(t)$ defined in Section 2A. Under the risk neutral probability $q_{t_{k+1}}^{(\Delta)}$,

$$C_{t_k}^{(\Delta)} = \left(1 + \bar{r}_{t_k}^{(\Delta)}\Delta_n,\right)^{-1} E^Q\left[C_{t_{k+1}}^{(\Delta)}\right], \qquad S_{t_k}^{(\Delta)} = \left(1 + \bar{r}_{t_k}^{(\Delta)}\Delta_n,\right)^{-1} E^Q\left[S_{t_{k+1}}^{(\Delta)}\right],$$
$$Z_{t_k}^{(\Delta)} = \left(1 + \bar{r}_{t_k}^{(\Delta)}\Delta_n,\right)^{-1} E^Q\left[Z_{t_{k+1}}^{(\Delta)}\right], \qquad \frac{Z_{t_k}^{(\Delta)}}{S_{t_k}^{(\Delta)}} = E^Q\left[\frac{Z_{t_{k+1}}^{(\Delta)}}{S_{t_{k+1}}^{(\Delta)}}\right]. \tag{23}$$

The last equation in (23)[7], confirms that, under the risk-neutral probability $q_{t_{k+1}}^{(\Delta)}$, the price ratio $Z_{t_k}^{(\Delta)}/S_{t_k}^{(\Delta)}$ is a martingale (i.e. the price of $\mathcal{Z}$ is a martingale under $q_{t_{k+1}}^{(\Delta)}$ when the price of $\mathcal{S}$ is used as a numéraire).[8]

By Davydov and Rotar (2008) and Hu et al. (2020), each return tree in (17a), (17b) generates a $\mathcal{D}[0,T]$ – process, which converges weakly in $\mathcal{D}[0,T]$ as $n \uparrow \infty$, under some smoothness and boundedness conditions on the limits $\mu_t$ of $\mu_{t_k}^{(\Delta)}$, $\sigma_t$ of $\sigma_{t_k}^{(\Delta)}$, $\tilde{\mu}_t$ of $\tilde{\mu}_{t_k}^{(\Delta)}$, and $\tilde{\sigma}_t$ of $\tilde{\sigma}_{t_k}^{(\Delta)}$ to define the continuous diffusion price processes $S_t$ and $Z_t$, $t \in [0,T]$, obeying $dS_t = S_t R_t^{(S)}$ and $dZ_t = Z_t R_t^{(Z)}$ where $R_t^{(S)}$ and $R_t^{(Z)}$ are the cumulative-return processes (Duffie, 2001, p.106) on $t \in [0,T]$,

$$dR_t^{(S)} = \mu_R(t)dt + \sigma_R(t)dW_t, \qquad \mu_R(t) = \mu_t, \qquad \sigma_R(t) = \sigma_t, \tag{24a}$$

[7] Showing this requires consideration of the four possibilities, $Z_{t_{k+1}}^{(\Delta,\mathrm{u})}/S_{t_{k+1}}^{(\Delta,\mathrm{u})}$, $Z_{t_{k+1}}^{(\Delta,\mathrm{u})}/S_{t_{k+1}}^{(\Delta,\mathrm{d})}$, $Z_{t_{k+1}}^{(\Delta,\mathrm{d})}/S_{t_{k+1}}^{(\Delta,\mathrm{u})}$ and $Z_{t_{k+1}}^{(\Delta,\mathrm{d})}/S_{t_{k+1}}^{(\Delta,\mathrm{d})}$ under the risk-neutral probability.

[8] The situation is symmetric; the price of $\mathcal{S}$ is a martingale under $q_{t_{k+1}}^{(\Delta)}$ if the price of $\mathcal{Z}$ is used as the numéraire.

$$dR_t^{(Z)} = \tilde{\mu}_R(t)dt + \tilde{\sigma}_R(t)dW_t, \qquad \tilde{\mu}_R(t) = \tilde{\mu}_t, \qquad \tilde{\sigma}_R(t) = \tilde{\sigma}_t\,. \tag{24b}$$

Here $W_t,\ t \in [0,T]$ is a standard Brownian motion (Wiener process) generated by the random walk $\xi_1^{(\Delta)} + \cdots + \xi_k^{(\Delta)}$, $k = 1,\dots,n$, as $n \uparrow \infty$. Under the risk-neutral measure $Q$, (24a) and (24b) become

$$dR_t^{(S)} = \bar{r}(t)dt + \sigma_t dW_t\,, \qquad dR_t^{(Z)} = \bar{r}(t)dt + \tilde{\sigma}_t dW_t\,, \tag{24c}$$

where $\bar{r}(t)$ is the shadow riskless rate for the market.

## 3. Approach Two

We consider a stock $\mathcal{S}$, a riskless bank account $\mathcal{B}$, and utilize a binomial pricing tree with a discrete stock price process $S_{t_k}^{(\Delta)}$ given by (16a) and return process $r_{t_{k+1}}^{(\Delta)}$ given by (17a). With the mean and variance of $r_{t_{k+1}}^{(\Delta)}$ given by $\mathbb{E}\left[r_{t_{k+1}}^{(\Delta)}\right] = \mu_{t_k}^{(\Delta)}\Delta_n$ and $\mathrm{Var}\left[r_{t_{k+1}}^{(\Delta)}\right] = \left(\sigma_{t_k}^{(\Delta)}\right)^2\Delta_n$, then $U_{t_{k+1}}^{(\Delta)}$ and $D_{t_{k+1}}^{(\Delta)}$ are given by (21a). The continuum limit of (16a), (17a) is the continuous diffusion process determined by (24a).

Suppose the riskless rate for $t \in (t_k, t_{k+1}]$, $k = 0,1,\dots,n-1$, is defined as $r_t^{(f,\Delta)} = r_{t_{k+1}}^{(f,\Delta)}$, with $r_{t_0}^{(f,\Delta)} \equiv r_0^{(f)} = 0$ and $D_{t_{k+1}}^{(\Delta)} < r_{t_{k+1}}^{(f,\Delta)} < U_{t_{k+1}}^{(\Delta)}$ to guarantee no-arbitrage (Shreve, 2004, equation (1.1.2)). The discrete dynamics of the riskless bank account $\mathcal{B}$ is given by

$$\beta_{t_k}^{(\Delta)} = \beta_{t_0}^{(\Delta)}\left(1 + r_{t_1}^{(f,\Delta)}\right)\dots\left(1 + r_{t_k}^{(f,\Delta)}\right), \qquad k = 1,\dots,n, \qquad \beta_{t_0}^{(\Delta)} \equiv \beta_0 = 1\,. \tag{25}$$

For $t \in (t_k, t_{k+1}]$, $k = 0,1,\dots,n-1$, we set $\beta_t^{(\Delta)} = \beta_{t_{k+1}}^{(\Delta)}$, defining a piecewise smooth price process for the bank account as well.

We seek a risk-neutral tree that preserves the natural probabilities for upward and downturn stock price movements. Consider the pricing tree (16a) under a change of the values for $U_{t_{k+1}}^{(\Delta)}$ and $D_{t_{k+1}}^{(\Delta)}$ by the numéraire $\pi_{t_{k+1}}^{(\Delta)} > 0$,

$$S_{t_{k+1}}^{(\pi,\Delta)} = \begin{cases} S_{t_{k+1}}^{(\pi,\Delta,\mathrm{u})} = S_{t_k}^{(\pi,\Delta)}\left(1 + U_{t_{k+1}}^{(\Delta)}\,\pi_{t_{k+1}}^{(\Delta)}\right) & \mathrm{w.p.} \quad p_{t_{k+1}}^{(\Delta)} \\ S_{t_{k+1}}^{(\pi,\Delta,\mathrm{d})} = S_{t_k}^{(\pi,\Delta)}\left(1 + D_{t_{k+1}}^{(\Delta)}\,\pi_{t_{k+1}}^{(\Delta)}\right) & \mathrm{w.p.} \quad 1 - p_{t_{k+1}}^{(\Delta)} \end{cases}, \quad k = 0,\dots,n-1. \tag{26}$$

The (independent) arithmetic returns under the changed numéraire are

$$r_{t_{k+1}}^{(\pi,\Delta)}=\frac{S_{t_{k+1}}^{(\pi,\Delta)}-S_{t_k}^{(\pi,\Delta)}}{S_{t_k}^{(\pi,\Delta)}}=\begin{cases} r_{t_{k+1}}^{(\pi,\Delta,\mathrm{u})}=U_{t_{k+1}}^{(\Delta)}\pi_{t_{k+1}}^{(\Delta)} & \text{w. p.} \quad p_{t_{k+1}}^{(\Delta)} \\ r_{t_{k+1}}^{(\pi,\Delta,\mathrm{d})}=D_{t_{k+1}}^{(\Delta)}\pi_{t_{k+1}}^{(\Delta)} & \text{w. p.} \quad 1-p_{t_{k+1}}^{(\Delta)} \end{cases},$$

$$k=0,\dots,n-1, \qquad r_{t_0}^{(\pi,\Delta)}\equiv r_0=0\,. \tag{27}$$

Thus,

$$S_{t_{k+1}}^{(\pi,\Delta)}=S_0\left(1+r_{t_1}^{(\pi,\Delta)}\right)\dots\left(1+r_{t_{k+1}}^{(\pi,\Delta)}\right), \qquad k=0,\dots,n-1. \tag{28}$$

The main assumption we make is that, at $t_{k+1}$, $k=0,1,\dots,n-1$, the risky asset $\mathcal{S}$ priced at the value $S_{t_{k+1}}^{(\pi,\Delta)}$ given by (28) is available for trade by market agents. That is, at $t_{k+1}$ a market agent, observing the history of returns $r_{t_1}^{(\pi,\Delta)},\dots,r_{t_{k+1}}^{(\pi,\Delta)}$ and knowing the value $S_0$, prices $\mathcal{S}$ by (28) rather than by (16a).

We determine $\pi_{t_{k+1}}^{(\Delta)}$ by requiring,

$$S_{t_k}^{(\pi,\Delta)}=\mathbb{E}_{t_k}\left[\frac{S_{t_{k+1}}^{(\pi,\Delta)}}{1+r_{t_{k+1}}^{(f,\Delta)}}\right]=\mathbb{E}\left[\frac{S_{t_{k+1}}^{(\pi,\Delta)}}{1+r_{t_{k+1}}^{(f,\Delta)}}\middle|\mathcal{F}_{t_k}^{(\Delta)}\right], \tag{29}$$

where $\mathcal{F}_{t_k}^{(\Delta)}=\sigma\left(\xi_1^{(\Delta)},\dots,\xi_k^{(\Delta)}\right)$ is the $\sigma$-field generated by $\xi_1^{(\Delta)},\dots,\xi_k^{(\Delta)}$. Thus we consider the stochastic basis for the pricing tree to be given by (Shreve, 2004)

$$\left(\Omega^{(\Delta)},\ \mathbb{F}^{(\Delta)}=\left\{\mathcal{F}_{t_k}^{(\Delta)},\ k=1,\dots,n,\ \mathcal{F}_{t_0}^{(\Delta)}=\mathcal{F}_0^{(\Delta)}=\{\emptyset,\Omega\}\right\},\ \mathbb{P}^{(\Delta)}\right). \tag{30}$$

From (29), (26) and (21a)

$$\pi_{t_{k+1}}^{(\Delta)}=\frac{r_{t_{k+1}}^{(f,\Delta)}}{U_{t_{k+1}}^{(\Delta)}p_{t_{k+1}}^{(\Delta)}+D_{t_{k+1}}^{(\Delta)}\left(1-p_{t_{k+1}}^{(\Delta)}\right)}=\frac{r_{t_{k+1}}^{(f,\Delta)}}{\mu_{t_k}^{(\Delta)}\Delta_n}=\frac{r_{t_{k+1}}^{(f,\Delta,\mathrm{inst})}}{\mu_{t_k}^{(\Delta)}}, \tag{31}$$

where

$$r_{t_{k+1}}^{(f,\Delta,\mathrm{inst})}=r_{t_{k+1}}^{(f,\Delta)}/\Delta_n, \tag{32}$$

denotes the instantaneous riskless rate of return at the discrete time point $t_{k+1}$. Using (21a) and (31), (26) becomes

$$S_{t_{k+1}}^{(\pi,\Delta)}= \tag{33}$$

$$\begin{cases} S_{t_{k+1}}^{(\pi,\Delta,\mathrm{u})} = S_{t_k}^{(\pi,\Delta)}\left(1 + r_{t_{k+1}}^{(f,\Delta,\mathrm{inst})}\Delta_n + \sigma_{t_k}^{(\Delta)}\dfrac{r_{t_{k+1}}^{(f,\Delta,\mathrm{inst})}}{\mu_{t_k}^{(\Delta)}}\sqrt{\dfrac{1-p_{t_{k+1}}^{(\Delta)}}{p_{t_{k+1}}^{(\Delta)}}}\sqrt{\Delta_n}\right) & \mathrm{w.p.} \quad p_{t_{k+1}}^{(\Delta)}, \\ S_{t_{k+1}}^{(\pi,\Delta,\mathrm{d})} = S_{t_k}^{(\pi,\Delta)}\left(1 + r_{t_{k+1}}^{(f,\Delta,\mathrm{inst})}\Delta_n - \sigma_{t_k}^{(\Delta)}\dfrac{r_{t_{k+1}}^{(f,\Delta,\mathrm{inst})}}{\mu_{t_k}^{(\Delta)}}\sqrt{\dfrac{p_{t_{k+1}}^{(\Delta)}}{1-p_{t_{k+1}}^{(\Delta)}}}\sqrt{\Delta_n}\right) & \mathrm{w.p.} \quad 1-p_{t_{k+1}}^{(\Delta)} \end{cases}$$

The mean and variance of the arithmetic return $r_{t_{k+1}}^{(\pi,\Delta)}$ are

$$\mathbb{E}\left[r_{t_{k+1}}^{(\pi,\Delta)}\right] = r_{t_{k+1}}^{(f,\Delta,\mathrm{inst})}\Delta_n, \qquad \mathrm{Var}\left[r_{t_{k+1}}^{(\pi,\Delta)}\right] = \left(\frac{r_{t_{k+1}}^{(f,\Delta,\mathrm{inst})}}{\mu_{t_k}^{(\Delta)}}\sigma_{t_k}^{(\Delta)}\right)^2\Delta_n. \tag{34}$$

The pricing tree (33) generates a $\mathcal{D}[0,T]$ process, which converges weakly in $\mathcal{D}[0,T]$ to the continuous diffusion price process $S_t^{(\pi)}$, $t \in [0,T]$, obeying $dS_t^{(\pi)} = S_t^{(\pi)}R_t^{(\pi)}$, where $R_t^{(\pi)}$, $t \in [0,T]$, is the cumulative-return process satisfying

$$dR_t^{(\pi)} = \mu_R^{(\pi)}(t)dt + \sigma_R^{(\pi)}(t)dW_t, \quad \mu_R^{(\pi)}(t) = r_t^{(f)}, \quad \sigma_R(t) = \frac{r_t^{(f)}}{\mu_t}\sigma_t\,, \tag{35}$$

and $r_t^{(f)}$, $t \in [0,T]$, is the short rate determined as the limit of $r_t^{(f,\Delta,\mathrm{inst})}$ under some smoothness and boundedness conditions.

Consider an option $\mathcal{C}^{(\pi)}$, with terminal payoff, $C_T^{(\pi,\Delta)} = g(S_T^{(\pi,\Delta)})$ where $g(x), x > 0$, satisfies the usual regularity conditions (Duffie, 2001, Chapter 5). Assume that an agent, taking a short position in $\mathcal{C}^{(\pi,\Delta)}$, forms a self-financing replicating portfolio by trading the asset $\mathcal{S}^{(\pi)}$ and depositing or borrowing the proceeds in the riskless bank account $\mathcal{B}$. Being short in the option, the agent takes a position in $\mathcal{S}^{(\pi)}$ so that the portfolio $P_{t_k}^{(\pi,\Delta)} = a_{t_k}^{(\Delta)}S_{t_k}^{(\pi,\Delta)} - C_{t_k}^{(\pi,\Delta)}$ earns the riskless rate of return. That is,

$$P_{t_{k+1}}^{(\pi,\Delta,\mathrm{u})} = a_{t_k}^{(\Delta)}S_{t_{k+1}}^{(\pi,\Delta,\mathrm{u})} - C_{t_{k+1}}^{(\pi,\Delta,\mathrm{u})} = P_{t_{k+1}}^{(\pi,\Delta,\mathrm{d})} = a_{t_k}^{(\Delta)}S_{t_{k+1}}^{(\pi,\Delta,\mathrm{d})} - C_{t_{k+1}}^{(\pi,\Delta,\mathrm{d})}, \tag{36}$$

requiring

$$a_{t_k}^{(\Delta)} = \frac{C_{t_{k+1}}^{(\pi,\Delta,\mathrm{u})} - C_{t_{k+1}}^{(\pi,\Delta,\mathrm{d})}}{S_{t_k}^{(\pi,\Delta)}\sigma_k^{(\Delta)}r_{t_{k+1}}^{(f,\Delta)}}\mu_k^{(\Delta)}\sqrt{p_{t_{k+1}}^{(\Delta)}\left(1-p_{t_{k+1}}^{(\Delta)}\right)}\sqrt{\Delta_n}\,. \tag{37}$$

As

$$P_{t_k}^{(\pi,\Delta)} = a_{t_k}^{(\Delta)} S_{t_k}^{(\pi,\Delta)} - C_{t_k}^{(\pi,\Delta)} = \frac{1}{1 + r_{t_{k+1}}^{(f,\Delta)}} P_{t_{k+1}}^{(\pi,\Delta,\mathrm{u})}, \tag{38}$$

it follows that

$$C_{t_k}^{(\pi,\Delta)} = \frac{1}{1 + r_{t_{k+1}}^{(f,\Delta)}} \left( p_{t_{k+1}}^{(\Delta)} C_{t_k}^{(\pi,\Delta,\mathrm{u})} + \left(1 - p_{t_{k+1}}^{(\Delta)}\right) C_{t_{k+1}}^{(\pi,\Delta,\mathrm{d})} \right). \tag{39}$$

As desired, the riskless and the natural probabilities are the same.

In the continuous-time limit, the price of the option with terminal time $T$ is

$$C_t^{(\pi)} = E_t \left[ e^{-\int_t^T r_s^{(f)} ds} g\left(S_T^{(\pi)}\right) \right]. \tag{40}$$

In the valuation of $C_t^{(\pi)}$ over $t \in [0, T]$, the instantaneous mean return $\mu_t$, the riskless rate $r_t^{(f)}$ and the volatility $\sigma_t$ play a role, and the expectation $E_t[\cdot]$ is under the natural (historical, real) probability $\mathbb{P}$.

This market model can be studied directly in continuous time starting with the dynamics $S_t, t \geq 0$, of the stock $\mathcal{S}$ as a continuous diffusion process,

$$dS_t = \mu_t S_t dt + \sigma_t S_t dW_t, \qquad t \geq 0, \tag{41}$$

on the complete stochastic basis $(\Omega, \mathcal{F}, \mathbb{F} = \{\mathcal{F}_t = \sigma(W_u, u \leq t), t \geq 0\}, \mathbb{P})$. From (41),

$$S_t = S_0 \exp\left[ \int_0^t \left( \mu_u - \frac{1}{2}\sigma_u^2 \right) du + \int_0^t \sigma_u dW_u \right], \qquad S_0 > 0. \tag{42}$$

Let $\mathcal{B}$ be a riskless bank account with price dynamics $\beta_t = \beta_0 \exp\left(\int_0^t r_s^{(f)} ds\right)$, $t \geq 0$, $\beta_0 > 0$, $r_t^{(f)} < \mu_t$. Under the standard BSM formulation, a call option $\mathcal{C}$ with $\mathcal{S}$ as the underlying has a price process $C_t(S_t, t)$ that obeys the PDE

$$\frac{\partial C_t}{\partial t} + (\sigma_t S_t)^2 \frac{\partial^2 C_t}{(\partial S_t)^2} + r_t^{(f)} S_t \frac{\partial C_t}{\partial S_t} = r_t^{(f)} C_t, \qquad t \geq 0, \tag{43}$$

subject to the final payoff $C_T = \max(S_T - K, 0)$.

Consider, instead, the continuous diffusion process $S_t^{(\pi)}, t \geq 0$, on $(\Omega, \mathcal{F}, \mathbb{F}, \mathbb{P})$ defined by

$$\frac{dS_t^{(\pi)}}{S_t^{(\pi)}} = \frac{dS_t}{S_t} \pi_t = (\mu_t dt + \sigma_t dW_t) \pi_t, \tag{44}$$

where $\pi_t > 0, t \geq 0$, is a deflator that acts on the instantaneous cumulative return $dR_t = dS_t / S_t$ rather than on the price process $S_t$. We refer to $\pi_t$ as the *cumulative return deflator* and assume that it is $\mathbb{F}$-adapted with $\sup\{\pi_t + 1/\pi_t,\ t \geq 0\} < \infty, \mathbb{P}\ a.s.$ We refer to the price process $S_t^{(\pi)}$ as having deflated cumulative returns. Paralleling (31), choose $\pi_t = r_t^{(f)}/\mu_t$ so that the instantaneous drift of the cumulative return of $S_t^{(\pi)}$is the riskless rate $r_t^{(f)}$ for $t \in [0, T]$,

$$dR_t^{(\pi)} = \frac{dS_t^{(\pi)}}{S_t^{(\pi)}} = \mu_R^{(\pi)}(t)dt + \sigma_R^{(\pi)}(t)dW_t, \qquad t \in [0, T],$$
$$\mu_R^{(\pi)}(t) = r_t^{(f)}, \qquad \sigma_R^{(\pi)}(t) = \frac{r_t^{(f)}}{\mu_t}\sigma_t, \tag{45}$$

exactly as in (35). From (45), for $t \in [0, T]$,

$$S_t^{(\pi)} = S_0^{(\pi)} \exp\left[\int_0^t \left(r_u^{(f)} - \frac{1}{2}\left(\sigma_R(u)\right)^2\right) du + \int_0^t \sigma_R(u) dW_u\right], \qquad S_0^{(\pi)} \equiv S_0 > 0\,. \tag{46}$$

Assuming a risky asset $\mathcal{S}^{(\pi)}$ with price process $S_t^{(\pi)}, t \geq 0$, and cumulative return process $R_t^{(\pi)}, t \geq 0$, is available for trade, then the market $\left(\mathcal{S}^{(\pi)}, \mathcal{B}\right)$ is arbitrage-free and complete. Furthermore, $S_t^{(\pi)}/\beta_t$ is a martingale on $(\Omega, \mathcal{F}, \mathbb{F}, \mathbb{P})$. Thus, the price of the option $\mathcal{C}^{(\pi)}$ with terminal time $T$ is given by (40). To determine the self-financing strategy replicating the option price $C_t^{(\pi)}\left(S_t^{(\pi)}, t\right)$, $t \geq 0$, in (40), consider the portfolio $P_t^{(\pi)} = a_t S_t^{(\pi)} + b_t \beta_t = C_t^{(\pi)}$ with $dP_t^{(\pi)} = a_t dS_t^{(\pi)} + b_t d\beta_t = dC_t^{(\pi)}$. Then the standard no-arbitrage arguments lead to

$$a_t = \frac{\partial C_t^{(\pi)}}{\partial S_t^{(\pi)}}, \qquad b_t = \frac{1}{\beta_t}\left(C_t^{(\pi)} - \frac{\partial C_t^{(\pi)}}{\partial S_t^{(\pi)}} S_t^{(\pi)}\right). \tag{47}$$

The corresponding BSM equation is then

$$\frac{\partial C_t^{(\pi)}}{\partial t} + \left(\sigma_R(t) S_t^{(\pi)}\right)^2 \frac{\partial^2 C_t^{(\pi)}}{\left(\partial S_t^{(\pi)}\right)^2} + r_t^{(f)} S_t^{(\pi)} \frac{\partial C_t^{(\pi)}}{\partial S_t^{(\pi)}} = r_t^{(f)} C_t^{(\pi)}, \qquad t \geq 0. \tag{48}$$

Assuming that $\mathcal{C}^{(\pi)}$ is a call option on $\mathcal{S}^{(\pi)}$ with price process $C_t^{(\pi)}, t \in [0, T]$ and final payoff $C_T^{(\pi)} = \max(S_T^{(\pi)} - K, 0)$, we obtain the classical BSM solution for the option.

We note that the option prices $C_t$ obtained from (43) and $C_t^{(\pi)}$ obtained from (48) are identical.

## 4. Specific Examples of Approach One

*A. $\mathcal{Z}$ has a Deflated Cumulative Return Process*

We return to approach one, considering a specific market of two risky assets $\mathcal{S}, \mathcal{S}^{(\pi)}$ and an option $\mathcal{C}$ on $\mathcal{S}$, where stock $\mathcal{S}$ has the price process (1) and asset $\mathcal{S}^{(\pi)}$ is stock $\mathcal{S}$ valued under the deflated cumulative return price process (46). We seek the equivalent martingale measure $\mathbb{Q}$ so that $Z_t^{(\pi)} \equiv S_t / S_t^{(\pi)}$ is a $\mathbb{Q}$-martingale. Under the change in probability measure $dW_t \ = dW_t^{\mathbb{Q}} - \theta_t dt$, with the market price of risk $\theta_t = \mu_t/\sigma_t - \sigma_t\, r_t^{(f)}/\mu_t$,

$$\begin{aligned} dZ_t^{(\pi)} &= Z_t^{(\pi)}\big(\sigma_t - \sigma_R(t)\big)dW_t^{\mathbb{Q}}, \\ dS_t &= S_t\big[\sigma_t\sigma_R(t)dt + \sigma_t dW_t^{\mathbb{Q}}\big]\,, \\ dS_t^{(\pi)} &= S_t^{(\pi)}\left[\big(\sigma_R(t)\big)^2 dt + \sigma_R(t)dW_t^{\mathbb{Q}}\right]. \end{aligned} \tag{49}$$

In the complete market $\big(\mathcal{S}, \mathcal{S}^{(\pi)}, \mathcal{C}\big)$ with price process $C_t, t \geq 0$, terminal time $T > 0$ and terminal payoff $C_T = g(S_T)$, $C_t/S_t^{(\pi)}$ is martingale under $\mathbb{Q}$ for $t \in [0, T]$ and

$$C_t = E_t^{\mathbb{Q}}\left(\frac{S_t^{(\pi)}}{S_T^{(\pi)}}\, g\Big(S_T^{(\pi)}\Big)\right). \tag{50}$$

From (49),

$$\begin{aligned} \frac{S_t^{(\pi)}}{S_T^{(\pi)}} &= \exp\left[-\int_t^T \frac{1}{2}\big(\sigma_R(u)\big)^2 du - \int_t^T \sigma_R(u)dW_u^{\mathbb{Q}}\right], \\ S_T &= S_t \exp\left[\int_t^T \left(\sigma_R(u)\sigma_u - \frac{1}{2}\sigma_u^2\right) du + \int_t^T \sigma_u dW_u^{\mathbb{Q}}\right]. \end{aligned} \tag{51}$$

Thus (48) becomes

$$\begin{aligned} C_t = E_t^{\mathbb{Q}}\Bigg[ &\exp\left\{-\int_t^T \frac{1}{2}\big(\sigma_R(u)\big)^2 du - \int_t^T \sigma_R(u)dW_u^{\mathbb{Q}}\right\} \\ &\times g\left(S_t \exp\left\{\int_t^T \left(\sigma_R(u)\sigma_u - \frac{1}{2}\sigma_u^2\right) du + \int_t^T \sigma_u dW_u^{\mathbb{Q}}\right\}\right)\Bigg]. \end{aligned} \tag{52}$$

This approach becomes an effective option pricing strategy provided the new (companion) asset type $\mathcal{S}^{(\pi)}$ becomes available for trade in the market. This new asset type is priced via the deflated cumulative return process.

### *B. $\mathcal{Z}$ is a Perpetual Derivative of $\mathcal{S}$*

We consider a second special case where the risky asset $\mathcal{Z}$ is a perpetual derivative of the asset $\mathcal{S}$.

B.1. Pricing the Perpetual Derivative

Consider a market containing stock $\mathcal{S}$ having price dynamics $S(t)$ as given in (1) and a riskless bank account $\mathcal{B}$ having price dynamics

$$dB_t = r_t^{(f)} B_t dt \, . \tag{53}$$

Let $\mathcal{D}$ denote a perpetual derivative of $\mathcal{S}$ having price $g(t, S_t)$ governed by the Itô process

$$dg = \left( \frac{\partial g}{\partial t} + \mu_t S_t \frac{\partial g}{\partial S_t} + \frac{\sigma_t^2 \, S_t^2}{2} \frac{\partial^2 g}{\partial {S_t}^2} \right) dt + \sigma_t S_t \frac{\partial g}{\partial S_t} dW_t \, . \tag{54}$$

To price $\mathcal{D}$, form the replicating portfolio $\pi^{(\mathcal{D})}(t) = a_t S_t + b_t B_t - g(t, S_t)$. The BSM PDE for the price dynamics of $\mathcal{D}$ is

$$r_t^{(f)} dg = \frac{\partial g}{\partial t} + r_t^{(f)} S_t \frac{\partial g}{\partial S_t} + \frac{\sigma_t^2 \, S_t^2}{2} \frac{\partial^2 g}{\partial {S_t}^2} , \tag{55}$$

with initial data $g(0, S_0)$. We investigate separable solutions to (55) of the form

$$g(t, S_t) = {S_t}^{\gamma} f(t) + h(t) \, , \tag{56}$$

where $\gamma$ is an arbitrary constant. From (55) we have the relation

$${S_t}^{\gamma} \left[ \eta(t) f(t) - \frac{df(t)}{dt} \right] = \frac{dh(t)}{dt} - r_t^{(f)} h(t) \, , \qquad \eta(t) \equiv (1 - \gamma) \left( r_t^{(f)} + \frac{\sigma_t^2}{2} \gamma \right) . \tag{57}$$

For (57) to hold, we require the coefficient of the stochastic term $S^{\gamma}(t)$ to vanish, leading to the solutions

$$f(t) = f(0) \exp\left( \int_0^t \eta(u) du \right) , \qquad h(t) = h(0) \exp\left( \int_0^t r_u^{(f)} du \right) . \tag{58}$$

The perpetual derivative price is then

$$g(t, S_t) = {S_t}^{\gamma} \exp\left( \int_0^t \eta(u) du \right) + \; h(0) \exp\left( \int_0^t r_u^{(f)} du \right) , \tag{59}$$

where we have set $f(0) = 1$ so that $g(0, S_0) = {S_0}^{\gamma} + h(0)$. The second term on the right-hand-side of (59) adds a deterministic, risk-free cumulative return to the perpetual derivative price. We, therefore, choose to concentrate on solutions where $h(0) = 0$. With the choice $h(0) = 0$, the dynamics (54) of the perpetual derivative is given by

$$\begin{aligned} dg(t, S_t) &= \tilde{\mu}(t) g(t, S_t) dt + \tilde{\sigma}(t) g(t, S_t) dW_t , \\ \tilde{\mu}(t) &= (1 - \gamma) r_t^{(f)} + \gamma \mu_t , \qquad \tilde{\sigma}(t) = \gamma \sigma_t \, . \end{aligned} \tag{60}$$

B.2. Examining the Perpetual Derivative

To understand the behavior of the perpetual derivative, we first consider the case in which the risk-free rate $r_t^{(f)}$, the drift $\mu_t$ and the volatility $\sigma_t$ have the respective time-independent values, $r^{(f)}$, $\mu$ and $\sigma$. From (59) (again with the choice $h(0) = 0$), the derivative price is

$$g(t, S_t) = {S_t}^{\gamma} e^{\xi(\gamma)\sigma^2 t/2}, \qquad \xi(\gamma) = (1-\gamma)(\delta+\gamma), \qquad \delta \equiv \frac{2r^{(f)}}{\sigma^2}. \tag{61}$$

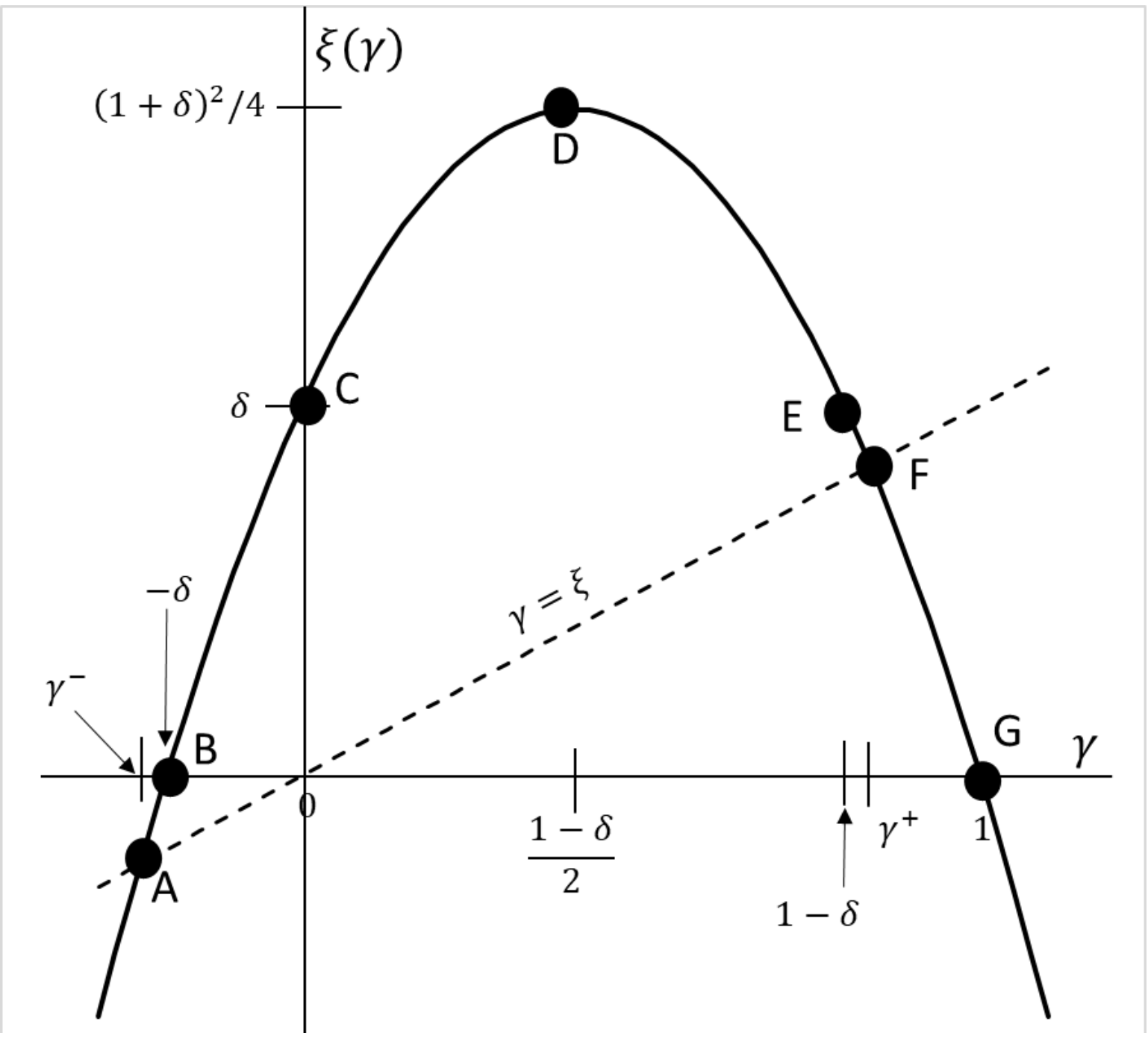

**Figure 1.** A sketch of the curve $\xi(\gamma)$ for the case in which the risk-free rate $r^{(f)}$ and the volatility $\sigma$ are time independent.

Fig. 1 sketches the dependence of $\xi(\gamma)$ on $\gamma$. Points labeled A through G represent specific solutions of interest. Solutions A and F correspond to the two values $\gamma^{\pm} = \left(-\delta(t) \pm \sqrt{\delta^2(t) + 4\delta(t)}\right)/2$ such that $\gamma^{\pm} = \xi(\gamma^{\pm})$ resulting in derivative prices $g(S_t) = {S_t}^{\gamma^{\pm}} e^{\gamma^{\pm}\sigma^2 t/2}$. Solutions B and G correspond to the case $\xi(\gamma) = 0$; solution B has the price $g(S_t) = {S_t}^{-\delta}$ while solution G with $g(S_t) = S_t$ corresponds to the stock price. Solutions C and E correspond to the case $\xi(\gamma) = \delta$; solution C having the price $g(S_t) = e^{\delta\sigma^2 t/2} = e^{rt}$ corresponding to a bank account, while solution E has price $g(S_t) = {S_t}^{1-\delta} e^{rt}$. Solution D corresponds to the maximum possible value of $\xi(\gamma) = (1+\delta)^2/4$ having the price $g(S_t) = {S_t}^{1-\delta} e^{(1+\delta)^2\sigma^2 t/8}$.

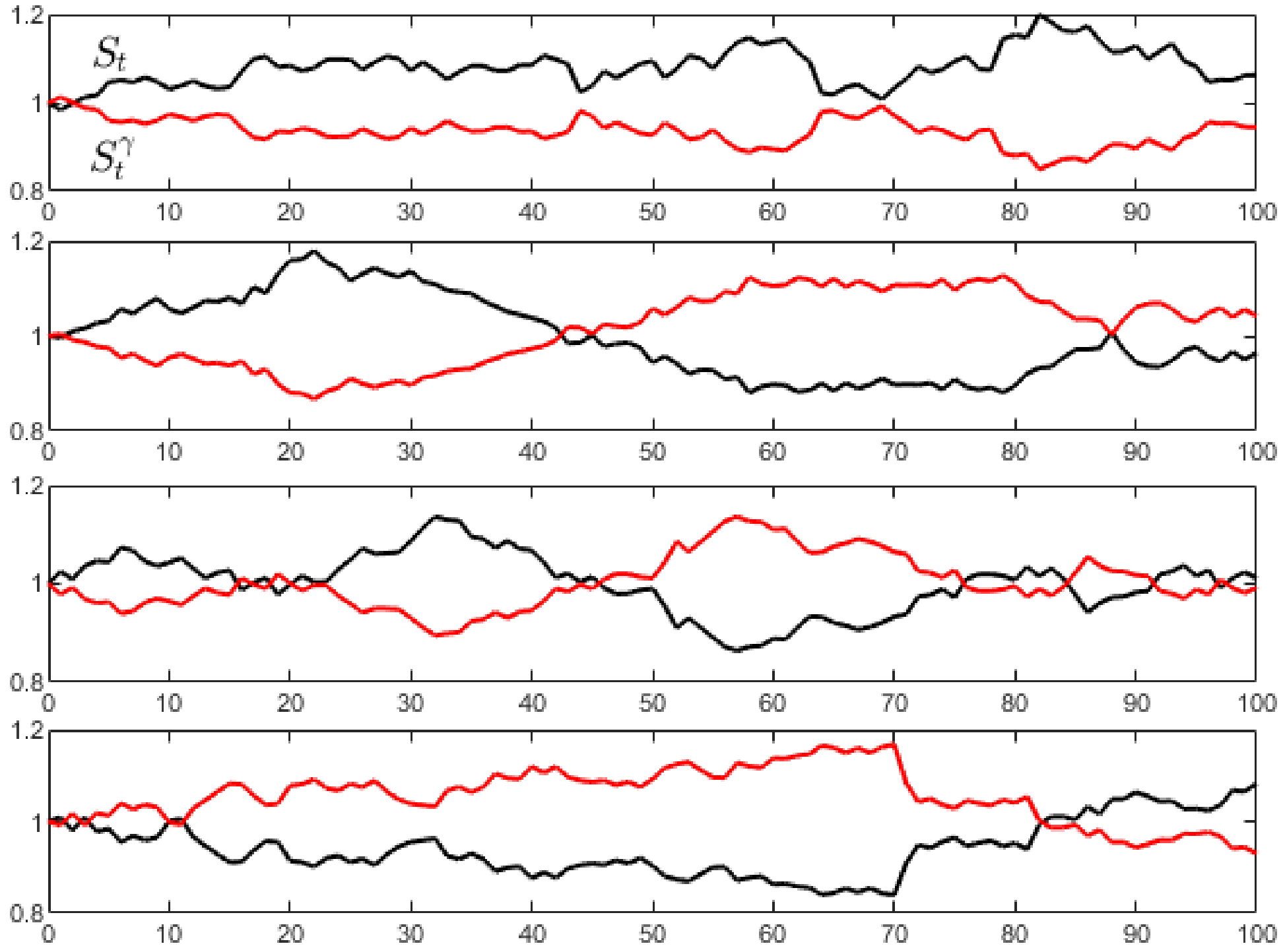

**Figure 2.** Example discrete paths of $S_t$ and $S_t^\gamma$, $\gamma = -\delta$, for the parameters given in (63).

Of particular interest is the choice $\gamma = -\delta = -2r^{(f)}/\sigma^2$ corresponding to solution B in Fig. 1, giving $g(S_t) = {S_t}^{-\delta}$. From (61) and (60), $g(S_t) = {S_t}^{-\delta}$ having the dynamics

$$d{S_t}^{-\delta} = {S_t}^{-\delta} \left((1+\delta)r^{(f)} - \delta\mu\right) dt - {S_t}^{-\delta}\delta\sigma dW_t \,. \quad (62)$$

Fig. 2 shows example discrete paths of prices $S_t$ and $S_t^{-\delta}$, (with $S_0 = S_0^{-\delta} = 1$) using the daily values[9] $(dt = 1)$

$$\mu = 4.38 \cdot 10^{-4}, \quad \sigma = 1.935 \cdot 10^{-2}, \qquad r^{(f)} = 1.635 \cdot 10^{-4} \quad (63)$$
$$\rightarrow \delta = 0.87332, \qquad \tilde{\mu} = -7.62 \cdot 10^{-5} \,.$$

The price of the stock and its derivative move "orthogonal" to each other.

[9] Based upon MSFT data for 08/29/2023 using a 512-day historical window and the 10-year Treasury yield for $r^{(f)}$.

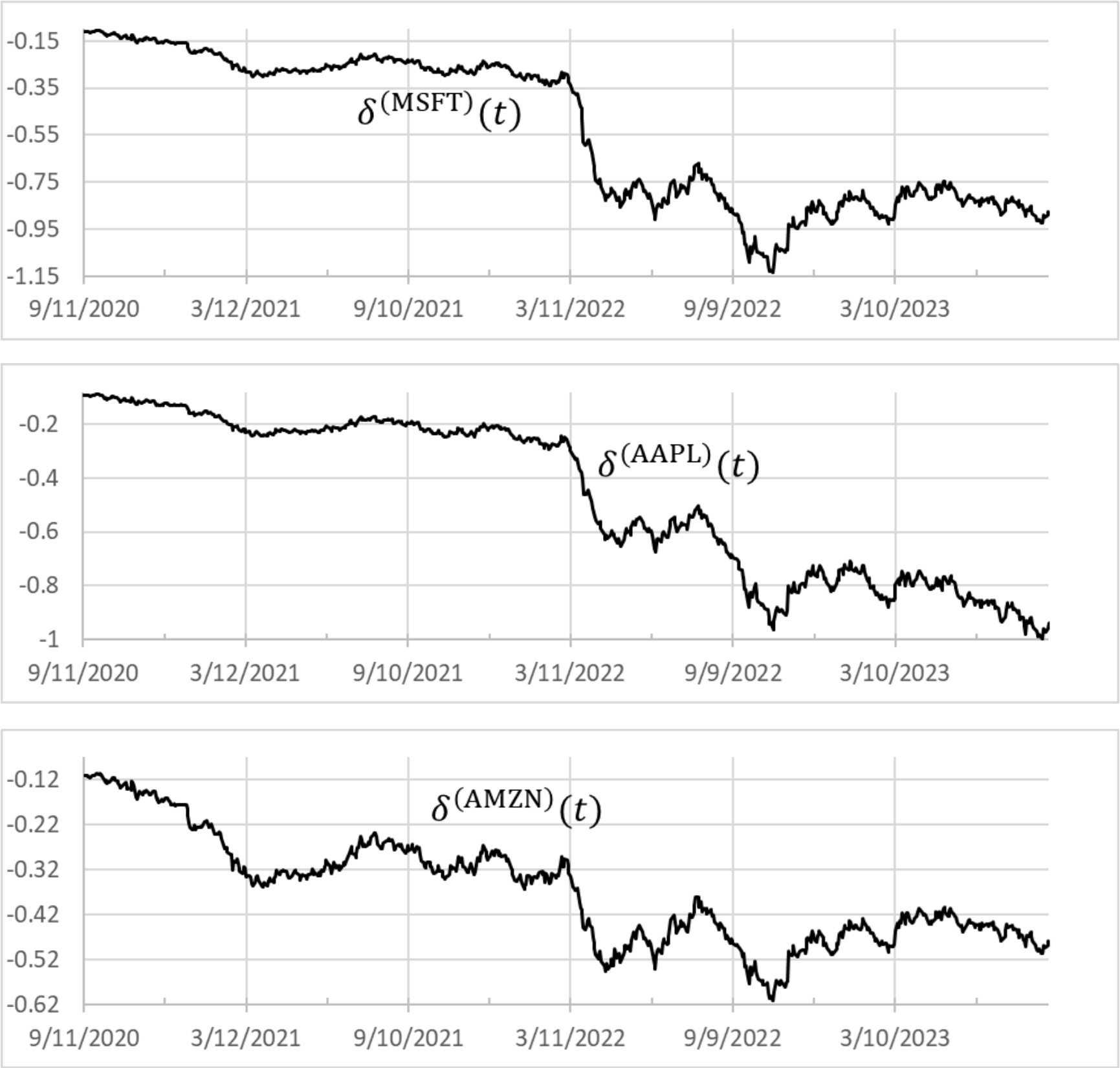

**Figure 3.** Estimated daily values of $\delta(t)$ for MSFT, AAPL and AMZN.

For the general case in which $r_t^{(f)}$, $\mu_t$ and $\sigma_t$ are time-dependent, a perpetual derivative is determined by fixing a choice of $\gamma$. As time progresses, the function $\xi(t;\gamma)$ will retain the general concave form shown in Fig. 1, but will scale and shift in a continuous manner, so that $\int_0^t \eta(u)du = 0.5\int_0^t \sigma_u^2\, \xi(u;\gamma)du$ integrates over changing values. Of course, one could hope that $\delta(t) = -2r_t^{(f)}/\sigma_t^2$ remains relatively constant over time, in which case a choice $\gamma \approx \delta(t)$ would produce a derivative whose price process is approximately $g(S_t) \approx S_t{}^{-\delta}$. Fig. 3, which plots estimated daily values[10] of $\delta(t)$ for MSFT, AAPL and AMZN over a four-year period, illustrates the limits of such an assumption.

[10] Based upon a 512-day moving window computation for $\sigma_t(t)$ and using 10-year U.W. treasury yields for $r_t^{(f)}$.

B.3. Option Pricing with the Perpetual Derivative

**Continuum model.** From approach 1, Section 2A, the market consists of stock $\mathcal{S}$, its perpetual derivative $\mathcal{S}^{(\gamma)}$(which plays the role of $\mathcal{Z}$ in Section 2A), and an option $\mathcal{C}$. The price dynamics of $\mathcal{S}$ and $\mathcal{S}^{(\gamma)}$ are given, respectively, by (1) and (60) (i.e., adopting the choice $h(0) = 0$). The price process for $\mathcal{C}$ is given by the LR PDE (4). Condition (5) leads to

$$\bar{r}(t) = r_t^{(f)}, \tag{64}$$

self-consistently identifying the shadow riskless rate $\bar{r}(t) = [\mu(t)\tilde{\sigma}(t) - \tilde{\mu}(t)\sigma(t)]/\Delta\sigma(t)$ as the risk-free rate in this market. The call option price $C(t, S_t, g_t; K, T)$ is given by (11). Assuming time-independent coefficients $\mu, \sigma, \bar{r}$ and the special case $g(S_t) = S_t{}^{-\delta}$ having dynamics driven by (62), the call option price is given by (15). However, the evaluation of $d = -y^*$ in (14) reduces to an equation of the form

$$\eta A(\bar{r}, \sigma)x + (1-\eta)B(\bar{r}, \sigma)x^{-\delta} = K, \qquad x \equiv S_t \exp\left(\sigma\sqrt{T-t}\, y^*\right), \tag{65}$$

which is still analytically intractable.

**Binomial model.** Applying the binomial tree pricing model of Section 2B to this market $\mathcal{S}, \mathcal{S}^{(\gamma)}, \mathcal{C}$, the risk-neutral probability and cumulative return (22) become

$$q_{t_{k+1}}^{(\Delta)} = p_{t_{k+1}}^{(\Delta)} - \frac{\left(\mu_{t_k}^{(\Delta)} - \bar{r}_{t_k}^{(\Delta)}\right)}{\sigma_{t_k}^{(\Delta)}} \sqrt{p_{t_{k+1}}^{(\Delta)}\left(1 - p_{t_{k+1}}^{(\Delta)}\right)\Delta_n}, \qquad R_{t_{k+1}}^{(\Delta)} = 1 + \bar{r}_{t_k}^{(\Delta)}\Delta_n. \tag{66}$$

The term is $\left(\mu_{t_k}^{(\Delta)} - \bar{r}_{t_k}^{(\Delta)}\right)/\sigma_{t_k}^{(\Delta)}$ is the market price of risk. The risk neutral probability is independent of $\gamma$, although the component of the payoff function due to $\mathcal{S}^{(\gamma)}$ will depend on the value of $\gamma$.

**5. Discussion**

Hu, Shirvani, Stoyanov, et al., (2020, sections 2.1 and 2.2) discuss two option price discontinuity puzzles. The first is the fact that the classical BSM solution produces an option price that is independent of the instantaneous mean return $\mu$ of the underlying stock. The BSM model states that the option hedger (the trader taking the short position in an option contract) does not care whether the underlying price is rapidly increasing (or decreasing). While this may be reasonable for option pricing during "normal" trading days, it fails completely under black swan market conditions such as Black Monday. Ultimately, the BSM model can only be valid under assumed conditions of frictionless, continuous trading (which break down during market turmoil).[11]

The second discontinuity puzzle is that option prices produced by the seminal Cox-Ross-Rubinstein and Jarrow-Rudd binomial pricing models are not continuous in the limits $p \downarrow 0$ or $p \uparrow 1$, where $p$ is the probability for upward price movement in these models. Unlike the continuous case, where the discontinuity can be "explained away" by invoking frictionless, continuous trading, the discontinuity in these binomial models are present in each single (time) period.

In contrast, option prices computed via binomial models developed by Kim et al. (2016), Hu, Shirvani, Stoyanov, et al. (2020), Hu, Shirvani, Lindquist, et al. (2020), and Hu et al. (2024) retain continuous dependence on the value of $p$. In addition, these various binomial models have been developed so that the option price contains information on $\mu$ as well as other stylized facts of real stock return processes. Unfortunately, in all these binomial models, the dependence of the option price on the instantaneous mean return $\mu$ vanishes in the continuous time limit.

Under both approaches described in this work, the option price developed under the binomial version of the approach retains continuous dependence on the instantaneous mean return and upward price movement probability. Furthermore, examination of (22) and (39) under the limits $p_{t_{k+1}}^{(\Delta)} \downarrow 0$ and $p_{t_{k+1}}^{(\Delta)} \uparrow 1$ reveals that randomness is removed from the option price (it becomes a fixed income security) in contrast to the behavior of the option price under the BSM model.

Under the first approach, the mean instantaneous rate vanishes in the continuum limit. (See (24c).) *However, in the continuous time limit of the second approach, the price of the option does depend on* $\mu$ *(as well as on the riskless rate and the volatility* $-$ *see (35)), thereby resolving the discontinuity puzzle of continuous-time BSM option pricing.* Equation (35) presents an interesting resolution of this discontinuity puzzle. For fixed $\sigma_t$ and $r_t^{(f)}$, as $\mu_t \uparrow \infty$, $\sigma_R(t) \downarrow 0$ and the risk neutral dynamics for the option price simplify to discounting a fixed deterministic payoff at maturity, in contrast to the BSM model where the option price is based upon a discounted expectation of a randomly distributed price at maturity.

[11] The generalization of the fundamental theorem of asset pricing by Delbaen an Schachermayer (1994), which introduced the concept of "no free lunch with vanishing risk" and follow-up work, particularly by Jarrow, Protter and Sayit (2009), show that no arbitrage conditions can be achieved by replacing the BSM restrictive assumption of frictionless, continuous trading, with the requirement of minimal fixed time between successive trades.

The second approach provides two additional benefits:

1. The option trader works entirely in the natural world – obviating the need to find a risk neutral measure.
2. The approach produces the same option prices as the classical approach in which prices are computed under the risk neutral measure $\mathbb{Q}$.

The second approach and the two specific examples of the first approach require the introduction of new, potentially marketable asset types, specifically a perpetual derivative $\mathcal{S}^{(\gamma)}$ of a stock, and a stock $\mathcal{S}^{(\pi)}$ whose price process is driven by a deflated cumulative return. These two asset types are designed specifically for hedgers who don't have access to sovereign riskless rates or may be hesitant to utilize interbank rates such as LIBOR or SOFR. The prices for $\mathcal{S}^{(\gamma)}$ and $\mathcal{S}^{(\pi)}$ can be computed as needed; however, we suggest that there is an opportunity for exchanges, such as Cboe, to develop such products for large-cap stocks.

## Appendix

We derive the required partial derivatives of $C(S, Z; K, T)$ and $d(t, S, Z) = -y^*(t, S, Z)$ and confirm that the option price (15) obeys the LR PDE (4). We begin by rewriting (14) as

$$\eta S(t) + (1-\eta)Z(t)\exp[-(\Delta w)^2/2 + \Delta w d] = Ke^{-m}\exp[-w^2/2 + wd], \tag{A1}$$

and further simplify this to

$$F_1(S) + F_2(Z, \Delta w, d) = F_3(t, w, d). \tag{A2}$$

Using the shorthand notation

$$\widehat{F_1} \equiv F_1/D\,, \qquad \widehat{F_2} \equiv F_2/D\,, \qquad \widehat{F_3} = \widehat{F_1} + \widehat{F_2}, \quad D \equiv wF_1 + \widetilde{w}F_2\,, \tag{A3}$$

taking partials of (A2), wrt to $t, S, Z$, we derive the relations

$$\begin{aligned}
d_t &= -\bar{r}\big(\widehat{F_1} + \widehat{F_2}\big) + \frac{d-w}{2(T-t)} - \frac{\Delta w\widetilde{w}\widehat{F_2}}{2(T-t)}\,,\\
Sd_S &= \widehat{F_1}, \qquad\qquad Zd_Z = \widehat{F_2},\\
S^2 d_{SS} &= \widehat{F_1}^{\,2}\big[\Delta w^2\widehat{F_2} - w^2\big(\widehat{F_1} + \widehat{F_2}\big)\big]\,,\\
SZd_{SZ} &= -w\widetilde{w}\widehat{F_1}\widehat{F_2}\big(\widehat{F_1} + \widehat{F_2}\big)\,,\\
Z^2 d_{ZZ} &= \big(\widehat{F_2}\big)^2\big[\Delta w^2\widehat{F_1} - \widetilde{w}^2\big(\widehat{F_1} + \widehat{F_2}\big)\big]\,.
\end{aligned} \tag{A4}$$

The derivatives of (15) appearing in the LR PDE are

$$C_t = \eta S\varphi(d)[d_t] + (1-\eta)Z\,\varphi(d-\Delta w)\left[d_t + \frac{\Delta w}{2(T-t)}\right]$$
$$-Ke^{-m}\Phi(d-w)[\bar{r}] - Ke^{-m}\varphi(d-w)\left[d_t + \frac{w}{2(T-t)}\right],$$
$$C_S = \eta\Phi(d) + \eta S\varphi(d)[d_S] + (1-\eta)Z\,\varphi(d-\Delta w)[d_S] - Ke^{-m}\varphi(d-w)[d_S],$$
$$C_Z = \eta S\varphi(d)[d_Z] + (1-\eta)\,\Phi(d-\Delta w) + (1-\eta)Z\,\varphi(d-\Delta w)[d_Z]$$
$$-Ke^{-m}\varphi(d-w)[d_Z],$$
$$C_{SS} = \eta\varphi(d)[2d_S - Sd(d_S)^2 + Sd_{SS}]$$
$$+ (1-\eta)Z\,\varphi(d-\Delta w)[d_{SS} - (d-\Delta w)(d_S)^2] \tag{A5}$$
$$-Ke^{-m}\varphi(d-w)[d_{SS} - (d-w)(d_S)^2],$$
$$C_{SZ} = \eta\varphi(d)[d_Z - Sdd_Sd_Z + Sd_{SZ}]$$
$$+ (1-\eta)\,\varphi(d-\Delta w)[d_S - Z(d-\Delta w)d_Sd_Z + Zd_{SZ}]$$
$$-Ke^{-m}\varphi(d-w)[d_{SZ} - (d-w)d_Sd_Z],$$
$$C_{ZZ} = \eta S\varphi(d)[d_{ZZ} - d(d_Z)^2]$$
$$+ (1-\eta)\,\varphi(d-\Delta w)[2d_Z - Z(d-\Delta w)(d_Z)^2 + Zd_{ZZ}]$$
$$-Ke^{-m}\varphi(d-w)[d_{ZZ} - (d-w)(d_Z)^2].$$

Adopting the notation

$$G_1 = \eta S\varphi(d),\quad G_2 = (1-\eta)\,\varphi(d-\Delta w),\quad G_3 = -Ke^{-m}\varphi(d-w),\quad G = G_1 + G_2 + G_3\,, \tag{A6}$$

the terms in (A5) can be rearranged more succinctly,

$$C_t = -Ke^{-m}\Phi(d-w)[\bar{r}] + G[d_t] + \left[\frac{\Delta wG_2 + wG_3}{2(T-t)}\right],$$
$$SC_S = \eta S\Phi(d) + G[Sd_S],$$
$$ZC_Z = (1-\eta)\,\Phi(d-\Delta w) + G[Zd_Z]\,, \tag{A7}$$
$$S^2C_{SS} = G[S^2d_{SS}] + (\Delta wG_2 + wG_3 - dG)[Sd_S]^2 + 2G_1[Sd_S]\,,$$
$$SZC_{SZ} = G[SZd_{SZ}] + (\Delta wG_2 + wG_3 - dG)[Sd_SZd_Z] + G_1[Zd_Z] + G_2[Sd_S]$$
$$Z^2C_{ZZ} = G[Z^2d_{ZZ}] + (\Delta wG_2 + wG_3 - dG)[Zd_Z]^2 + 2G_2[Zd_Z]$$

From the definitions of $w$ and $\widetilde{w}$, we have the trivial observations

$$\sigma^2 = \frac{w^2}{T-t},\quad \sigma\tilde{\sigma} = \frac{w\widetilde{w}}{T-t},\quad \tilde{\sigma}^2 = \frac{\widetilde{w}^2}{T-t}\,. \tag{A8}$$

Substituting (15) and (A7) into the LR PDE produces

$$\bar{r}[\eta S_t\Phi(d) + (1-\eta)Z_t\,\Phi(d-\Delta w)\; - Ke^{-m}\Phi(d-w)]$$

$$= -\bar{r}Ke^{-m}\Phi(d-w) + G[d_t] + \left[\frac{\Delta w G_2 + wG_3}{2(T-t)}\right]$$

$$+\bar{r}\eta S\Phi(d) + \bar{r}(1-\eta)\,\Phi(d-\Delta w) + \bar{r}G[Sd_S + Zd_Z]$$

$$+\frac{1}{2(T-t)}(w^2S^2C_{SS} + 2w\tilde{w}SZC_{ZS} + \tilde{w}^2Z^2C_{ZZ})\,. \qquad \text{(A9)}$$

Cancelling common terms involving the cumulative distribution $\Phi(\cdot)$ reduces (A9) to

$$0 = G[d_t] + \left[\frac{\Delta w G_2 + wG_3}{2(T-t)}\right] + \bar{r}G[Sd_S + Zd_Z]$$

$$+\frac{1}{2(T-t)}(w^2S^2C_{SS} + 2w\tilde{w}SZC_{ZS} + \tilde{w}^2Z^2C_{ZZ})\,. \qquad \text{(A10)}$$

Substituting the $d_t$, $Sd_S$ and $Zd_Z$ terms from (A4) produces

$$0 = G\left[-\bar{r}\left(\widehat{F_1} + \widehat{F_2}\right) + \frac{d-w}{2(T-t)} - \frac{\Delta w\tilde{w}\widehat{F_2}}{2(T-t)}\right] + \left[\frac{\Delta w G_2 + wG_3}{2(T-t)}\right] + \bar{r}G\left[\widehat{F_1} + \widehat{F_2}\right]$$

$$+\frac{1}{2(T-t)}(w^2S^2C_{SS} + 2w\tilde{w}SZC_{ZS} + \tilde{w}^2Z^2C_{ZZ})\,. \qquad \text{(A11)}$$

The terms in $\bar{r}$ cancel, reducing (A11) to

$$0 = Gd - wG_1 - \tilde{w}G_2 + G\widehat{F_2}(\tilde{w}^2 - w\tilde{w}) + w^2S^2C_{SS} + 2w\tilde{w}SZC_{ZS} + \tilde{w}^2Z^2C_{ZZ}\,. \qquad \text{(A12)}$$

We concentrate on the last three terms in (A12). From (A7)

$$w^2S^2C_{SS} + 2w\tilde{w}SZC_{ZS} + \tilde{w}^2Z^2C_{ZZ}$$

$$= G[w^2S^2d_{SS} + 2w\tilde{w}SZd_{ZS} + \tilde{w}^2Z^2d_{ZZ}]$$

$$+[wG_2 + wG_3 - \tilde{w}G_2 - dG][w^2(Sd_S)^2 + 2w\tilde{w}(Sd_SZd_Z)$$

$$+\tilde{w}^2(Zd_Z)^2]$$

$$+2[wG_1(wSd_S + \tilde{w}Zd_Z) + \tilde{w}G_2(wSd_S + \tilde{w}Zd_Z)]\,. \qquad \text{(A13)}$$

Using equations (A4) and the identity

$$w\widehat{F_1} + \tilde{w}\widehat{F_2} = 1, \qquad \text{(A14)}$$

we derive

$$w^2S^2d_{SS} + 2w\tilde{w}SZd_{ZS} + \tilde{w}^2Z^2d_{ZZ} = -\left(w^2\widehat{F_1} + \tilde{w}^2\widehat{F_2}\right),$$

$$w^2(Sd_S)^2 + 2w\tilde{w}(Sd_SZd_Z) + \tilde{w}^2(Zd_Z)^2 = \left(w\widehat{F_1} + \tilde{w}\widehat{F_2}\right)^2 = 1, \qquad \text{(A15)}$$

$$wG_1(wSd_S + \tilde{w}Zd_Z) + w_ZG_2(wSd_S + \tilde{w}Zd_Z) = wG_1 + w_ZG_2.$$

Putting these results in (A13) produces

$$w^2S^2C_{SS} + 2w\widetilde{w}SZC_{ZS} + \widetilde{w}^2Z^2C_{ZZ}$$
$$= -G\left[w^2\widehat{F_1} + \widetilde{w}^2\widehat{F_2}\right] + wG_2 + wG_3 - \widetilde{w}G_2 - dG + 2(wG_1 + \widetilde{w}G_2)\,. \quad \text{(A16)}$$

Substituting (A16) into (A12) gives

$$\begin{aligned} 0 &= Gd - wG_1 - \widetilde{w}G_2 + G\widehat{F_2}(\widetilde{w}^2 - w\widetilde{w}) \\ &\quad -G\left[w^2\widehat{F_1} + \widetilde{w}^2\widehat{F_2}\right] + wG_2 + wG_3 - \widetilde{w}G_2 - dG + 2wG_1 + 2\widetilde{w}G_2\,. \end{aligned} \quad \text{(A17)}$$

Canceling common terms produces

$$w(G_1 + G_2 + G_3) - wG\left(w\widehat{F_1} + \widetilde{w}\widehat{F_2}\right) = wG - wG(1) = 0\,, \quad \text{(A18)}$$

confirming that the call option price (15) satisfies the LR PDE.